\documentclass[seceq]{ptptex}

\usepackage{bm}
\newcommand{\be}{\begin{equation}}
\newcommand{\ee}{\end{equation}}
\newcommand{\bea}{\begin{eqnarray}}
\newcommand{\eea}{\end{eqnarray}}

\newcommand{\rr}{{\bf r}}

\newcommand{\qq}{{\bf q}}

\newcommand{\jj}{{\bf j}}
\newcommand{\vv}{{\bf v}}
\newcommand{\kk}{{\bf k}}

\newcommand{\rme}{{\rm e}}
\newcommand{\rmi}{{\rm i}}

\title{Superfluid dynamics in neutron star crusts
}

\author{
C.\ J.\  \textsc{Pethick}$^1$, N.\ \textsc{Chamel}$^2$, Sanjay \textsc{Reddy}$^3$}%
\inst{
$^1$\mbox{The Niels Bohr International Academy, The
Niels Bohr Institute,}\\
\mbox{Blegdamsvej 17, DK-2100 Copenhagen \O,
 Denmark}\\
\mbox{NORDITA, Roslagstullsbacken 23, SE-10691 Stockholm, Sweden} \\
$^2$\mbox{Institut d'Astronomie et d'Astrophysique, CP-226, Universit\'e Libre de Bruxelles,}
\mbox{
	BE-1050 Brussels, Belgium}
	\\
$^3$\mbox{T16, MS B283, Los Alamos National Laboratory, Los Alamos NM 87545, U.\,S.\,A.}}	
	
\abst{
A simple description of superfluid hydrodynamics in the inner crust of a neutron star is given.  Particular attention is paid to the effect of the lattice of nuclei on the properties of the superfluid neutrons, and the effects of entrainment, the fact that some fraction of the neutrons are locked to the motion of the protons in nuclei.
}

\begin{document}

\maketitle

\section{Introduction}
The outer layers of a neutron star at densities below nuclear saturation density make up only a small part of the star, but they are important because processes that occur there are more accessible to observation than are those in the inner parts of stars.  Among phenomena for which the physics of the outer layers of the star is crucial one may mention the thermal evolution of neutron stars.  This includes the recent suggestion that heat transport by superfluid neutrons in the crust could be a significant contributor to thermal transport \cite{aguilera}.   Another example of the importance of  the crust is the proposal that quasiperiodic oscillations seen in x-ray afterglows in giant flares on neutron stars may be attributed to excitation of torsional modes of oscillation of the crust \cite{duncan,strohmeyer,andersson}.
In both these contexts, modes of oscillation in the crust play a key role, and the purpose of this article is to present a simple approach to calculating their properties.  

The focus of this article is the inner crust of neutron stars, where superfluid neutrons coexist with a lattice of ions.
There is an extensive literature on the physics of this region (e.g. \cite{lrr}). However it has been only recently pointed 
out that an important role is played by ``entrainment'' of neutrons by the protons in nuclei~\cite{CCH2}. Such
mutual entrainment effects have been studied for a long time in the liquid core of neutron stars (see e.g. ~\cite{sauls89}). 
In this article, we begin by describing how to formulate the theory of superfluidity when there is a periodic lattice of ions.  We then present the hydrodynamic equations and show that, as a consequence of Galilean invariance, an appropriately defined superfluid density characterizes both the effective mass of nuclei and the strength of a vector coupling between lattice vibrations and sound waves in the neutron superfluid.  We derive the dispersion relation for the long-wavelength collective modes, which are generally a superposition of lattice vibrations and sound waves in the neutron superfluid.  
Finally,  we comment on the physical meaning of the neutron superfluid density.  

\section{Neutron superfluidity in the presence of a lattice of nuclei}

In a Fermi superfluid, the pairing amplitude at position $\rr$ is given by  $\langle \psi_{ \uparrow}(\rr) \psi_{ \downarrow}(\rr) \rangle$, where 
$\psi_{\sigma}(\rr)$ is the annihilation field operator for a neutron with spin $\sigma$ at position $\rr$, 
and $\langle\ldots\rangle$ denotes a quantum-mechanical or, at nonzero temperature, a thermal average. 
For a superfluid at rest, the phase $2\varphi({\bf r})$ of the pairing amplitude is independent of position, while for a superfluid in motion it depends on space. In a translationally invariant system with paired neutrons, such as neutron matter 
in the outer core of a neutron star,  the phase of the pairing amplitude varies smoothly in space, while in the presence of a crystal lattice, the phase has, in addition, oscillatory contributions with a period equal to the lattice spacing.  The question is know what quantity should one use to describe motions of the superfluid on length scales large compared with the lattice spacing.

Just as for single-electron wave functions in a periodic lattice, one may characterize states of 
the neutron superfluid in the crust by a wave vector $\kk$ such that the change of the phase $\varphi(\rr)$ when the spatial coordinate is displaced by a lattice vector ${\bf R}$ is given by $\kk\cdot {\bf R}$.   Alternatively, one may work in terms of a coarse-grained average phase, $\phi(\rr)$, which is the average of $\varphi$ over a region in the vicinity of $\rr$ with dimensions large compared with the lattice spacing but small compared with other length scales in the problem, in which case $\kk=\bm\nabla \phi$.  Similar physics occurs for atomic Bose--Einstein condensates in artificial crystal lattices created by standing-wave laser beams, so-called optical lattices \cite{machholm}.

\section{Hydrodynamic equations}  We now derive equations of motion for the superfluid and the crystal lattice in situations when the scale of spatial variations is large compared with the lattice spacing.  In addition, we shall require that the scale of spatial variations is large compared with the size of the ``neutron pair wave function'', which is given by the superfluid coherence length, $\xi\sim \hbar v_{Fn}/\Delta_n$, where  $v_{Fn}$ is the neutron Fermi velocity   and $\Delta_n$ the neutron superfluid energy gap.  We shall treat the nucleons as nonrelativistic, which is a good first approximation in neutron-star crusts.  For simplicity, we shall neglect the effects of the gravitational potential. Since this varies relatively little over the crust, it is a reasonable approximation to treat space as flat.

Hydrodynamic equations are an expression of conservation laws for momentum and the numbers of neutrons and protons.  Our treatment follows closely the standard discussion of the two-fluid model for liquid helium-4 \cite{LLHydro}.  For simplicity, we shall assume here that the temperature is zero and we shall neglect the external potential and inhomogeneity of the equilibrium density distribution.  We shall consider disturbances whose length scale is large compared with the screening length for electrons, in which case matter is electrically neutral locally, and the number density of electrons is equal to that of the protons.  Two of the variables with which we choose to work are two densities - the total neutron density $n_n$ and the proton density $n_p$.  This choice is particularly convenient, since these are the quantities that enter the conservation laws for particle number.  Alternatively, one could have chosen use appropriately defined densities of neutrons associated with nuclei and with the ``superfluid between nuclei''\cite{CCH2}.  This does not alter the final results,  but  the formalism is particularly straightforward when $n_n$ and $n_p$ are used.
The two other variables we employ describe the motion of the constituents.  We shall assume that the ions are sufficiently far apart that hopping of protons from one nucleus to another is absent, so we may describe the configuration of the ions in terms of the displacement ${\bm\xi}(\rr)$ of an ion from its equilibrium position.  Thus the velocity of the ions is given by ${\bf v}_p=\dot {\bm\xi}$.  For the superfluid, we characterize its motion by the wave vector $\kk$, or equivalently the ``neutron superfluid  velocity'', which we define as \cite{neutronnormal}
\be
 {\bf v}_n =\frac{\hbar \kk}{m},
 \label{v_n}
\ee
where $m$ is the nucleon mass (we neglect the difference between the neutron and proton masses).
The choice of the mass used in this expression is a matter of convention, since the important quantity is really the wave vector.
The first two conservation laws we consider are those for conservation of the number of neutrons and the number of protons.  In the case of protons, the continuity equation, 
when linearized, is
\be
\frac{\partial n_p}{\partial t} +n_p{\bm\nabla} \cdot \vv_p=0.
\label{conp2}
\ee
Throughout, we shall consider small departures of the system from equilibrium, and therefore we shall linearize the equations. The full non-linear equations can be found in Refs.~ \cite{CCH2,CartChach}.
We shall assume that the atomic number $Z$ of nuclei is a constant in the region under consideration, and therefore the proton density is given by $n_p=Z n_I$, where $n_I$ is the density of nuclei.
To write a conservation law for the neutron number, we need an expression for the current density of neutrons.  Consider first the case where the ions are at rest, and we denote quantities in this frame by a prime.  Phenomenologically, for small superfluid velocities  one may write the current density of neutrons as
\be
\jj'_n=n_n^s \vv'_n,
\ee
 where $n_n^s$ may be regarded as the density of superfluid neutrons.  Now let us consider the current in a frame moving with velocity $-\vv_p$.  By Galilean invariance, the current in the new frame is given by
 \be
 \jj_n=n_n^s\vv'_n +n_n\vv_p=n_n^s\vv_n +n_n^b\vv_p.
 \label{neutroncurrent}
 \ee
 Here $\vv_n=\vv'_n+\vv_p$ is the neutron superfluid velocity in the new frame.  This result follows from the fact that under the Galilean transformation, every single nucleon state acquires an extra phase factor $\exp(\rmi m \vv_p\cdot \rr_i/\hbar)$, $\rr_i$ being the nucleon coordinate, and therefore the phase $\varphi(\rr)$ is increased by $m \vv_p\cdot \rr/\hbar$ . The quantity $n_n^b=n_n-n_n^s$ may be regarded as the density of neutrons bound to nuclei.
 The continuity equation for neutrons is therefore 
 \be
\frac{\partial n_n}{\partial t} +n_n^s\bm\nabla \cdot \vv_n +n_n^b\bm\nabla \cdot \vv_p=0.
\label{conn}
\ee
The next equation is that for conservation of momentum, which in the absence of dissipation reads
\be
\frac{\partial {\bf g}}{\partial t} ={\bf f},
\label{momcons}
\ee
where $\bm f$ is the force per unit volume.    The momentum density may be calculated in a similar way to that in which we calculated the current above.     If effects of relativity on the motion of the neutrons are neglected, the momentum density is given by the current density, times the nucleon rest mass, and thus, in the frame in which the protons are at rest,
\be
\bm g'_n=mn_n^s \vv'_n.
\ee   
By making a Galilean transformation to a frame moving with velocity $-\vv_p$, one finds that the momentum density is changed by an amount $\rho \vv_p$, and therefore the momentum density is given by
\be
\bm g=\rho \vv_p +mn_n^s\vv_n'=\rho_p^{\rm eff}\vv_p+mn_n^s\vv_n.
\ee
Here $\rho$ is the total mass density, the energy density divided by $c^2$, and $\rho_n^{\rm eff}=\rho-mn_n^s$. If the electron contribution to $\rho$ is neglected, $\rho^{\rm eff}=m(n_p+n_n^s)$, and the momentum density is the nucleon current density times $m$:
\be
{\bf g}=mn_n^s \vv_n+m(n_p+n_n^b)\vv_p.
\ee 
The condition for momentum conservation, (\ref{momcons}), is therefore
\be
mn_n^s\frac{\partial {\vv_n}}{\partial t}+m(n_p+n_n^b)\frac{\partial {\vv_p}}{\partial t}  ={\bm f}.
\label{momcons2}
\ee
 When the rigidity of the crystal lattice is neglected, the system may be treated as coupled fluids and the force is given by 
\be
{\bm f}=-{\bm \nabla}p,
\ee
where $p$ is the pressure.  At zero temperature, changes in pressure are related to changes in the neutron and proton chemical potentials by the Gibbs--Duhem relation
\be
dp=n_nd\mu_n+n_pd\mu_p,
\ee
and the condition for momentum conservation becomes
\be
mn_n^s\frac{\partial {\vv_n}}{\partial t}+m(n_p+n_n^b)\frac{\partial {\vv_p}}{\partial t} +n_n\bm\nabla \mu_n+n_p\bm\nabla \mu_p=0.
\label{momconscont0}
\ee

When the protons form a rigid lattice of nuclei, it is necessary to include the effects of elasticity.  For an isotropic solid, the elastic contribution to the energy is given to second order in the strain tensor 
\be
u_{ij}=\frac12\left(\frac{\partial u_i}{\partial x_j}+\frac{\partial u_j}{\partial x_i}\right)
\ee
 by \cite{LLElastic}
\be
E_{\rm elastic}=\frac12\tilde  K u_{ii}^2 + S\left(u_{ij}-\frac{\delta_{ij}u_{kk}}3\right)^2.
\ee
where $\tilde K=\partial^2 E/\partial u_{ii}^2|_{n_n}$ and $S$ is the shear modulus.  
In a solid with lower symmetry, the expression for the elastic energy has more terms.  For crustal material with roughly spherical nuclei, the crystal symmetry is expected to be body-centered cubic.  However, it seems unlikely that the crust behaves as a perfect crystal, and one would expect it to be polycrystalline.  On length scales large compared with the size of a domain, the solid will behave as an isotropic solid.  For phases with liquid-crystal-like phases, the elastic properties are even more complicated, and the effects of imperfect ordering of nuclei has yet to be investigated \cite{potekhin}.

In a neutron star crust, there is also a contribution due to coupling of the neutrons to the strain field, and the contribution linear in the strain and linear in the changes in the neutron density has the form 
\be
E_{n,{\rm lattice}}=-L\delta n_n u_{ii},
\ee
which is the only term consistent with rotational invariance.  The trace $u_{ii}$ of the strain tensor is related to changes in the proton density by the equation
\be
u_{ii}=-\frac{\delta n_p}{n_p}.
\ee
 Thus one can see that 
 \be
 L=-\frac{\partial^2 E}{\partial n_n\partial u_{ii}}=n_p\frac{\partial \mu_n}{\partial n_p}.
 \ee
Similarly, one can show that $\tilde K=n_p^2\partial \mu_p/\partial n_p$.  For an elastic solid, the derivatives with respect to $n_p$ are to be evaluated with the shear strain held fixed.
For a solid consisting of a single component, the quantity $\tilde K$  would be the bulk modulus, whereas for the system under consideration here,  it is the contribution to the bulk modulus when the density of neutrons is held fixed. 
 The total bulk modulus is given by $K=-V\partial p/\partial V$, where $V$ is the volume of the system and the derivative is to be evaluated keeping the numbers of neutrons and protons fixed, and therefore
$K=\tilde K+2n_n L +n_n^2{\partial \mu_n}/{\partial n_n}$.
Quite generally, the quantity $\partial \mu_i/\partial n_j$ represents an effective interaction between densities of components $i$ and 
$j$. Thus the effective proton--proton interaction is $\tilde K/n_p^2$ and the neutron--proton one $L/n_p$.

With the effects of elasticity included, the force per unit volume is given by
\be
 f_i=-n_n \nabla_i \mu_n +\tilde K\nabla_i u_{jj}-L\nabla n_n +2S\nabla_ju_{ij},
\ee
and the condition for momentum conservation becomes
\be
mn_n^s\frac{\partial {v_{ni}}}{\partial t}+m(n_p+n_n^b)\frac{\partial v_{pi}}{\partial t} +n_n\nabla_i \mu_n-\tilde K\nabla_i u_{jj} +L\nabla_i n_n-2S\nabla_j u_{ij}=0.
\label{momconslattice0}
\ee

A fourth condition is needed to close the system of equations.  This is the Josephson relation, which gives the time evolution of the phase of the superfluid neutrons.  In a uniform system, the rate of change of the phase   
is given by
\be
\frac{\partial\phi} {\partial t}=-\frac{\mu_n}{\hbar},
\label{josephson}
\ee
and provided spatial variations are sufficiently slow, this condition will also apply locally.  The spatial gradient  of this relation gives an equation for the evolution of the superfluid velocity which, because of Eq.\ (\ref{v_n}), is
\be
\frac{\partial {\vv_n}}{\partial t} +\frac1m\bm\nabla\mu_n=0.
\ee
If one subtracts $n_n^s m$ times this equation from Eqs.\ (\ref{momconscont0}) and (\ref{momconslattice0})
one finds
\be
m(n_p+n_n^b)\frac{\partial {\vv_p}}{\partial t} +n_n^b\bm\nabla \mu_n+n_p\bm\nabla \mu_p=0,
\label{momconscont}
\ee
and 
\be
m(n_p+n_n^b)\frac{\partial v_{pi}}{\partial t} +n_n^b\nabla_i \mu_n+L\nabla_in_n -\tilde K\nabla_i u_{jj} -2S\nabla_ju_{ij}=0.
\label{momconslattice}
\ee
Equations (\ref{momconscont}) and (\ref {momconslattice}) show that nuclei behave as though they had a mass density $m((n_p+n_n^b))/n_I=Zm(1+n_n^b/n_p)$.  The quantity $n_n^b$ thus represents the number density of neutrons that should be regarded as being bound to protons, so far as the inertia of the nuclei is concerned.  

\section{Normal modes}

As an illustration we consider normal modes of oscillation of the system, which are obtained by solving the linearized hydrodynamic equations, treating velocities and deviations of the densities from their equilibrium values as small quantities.   We expand the deviations of the neutron and proton chemical potential by writing for the fluid case
\be
d\mu_n =\frac{\partial \mu_n}{\partial n_n} dn_n +\frac{\partial \mu_n}{\partial n_p} dn_p
\ee
and the corresponding equation obtained by interchanging the neutron and proton labels.
For the case of a solid, the neutron chemical potential is given by $\mu_n=\partial E/\partial n_n|_{u_{ij}}$, and therefore
\be
d\mu_n=\frac{\partial \mu_n}{\partial n_n} dn_n -Ldu_{ii}.
\ee 
The normal modes have the form of plane waves that vary in space and time as $\rme^{\rmi(\qq.\rr-\omega t)}$, where $\qq$ is the wave vector and $\omega$ the frequency.  In an isotropic medium, the normal modes may be separated into transverse and longitudinal ones. At long wavelengths, the normal modes all have a sound-like dispersion relation, with $\omega=vq$, $v$ being the mode velocity.  The transverse modes are the simpler, since the neutron superfluid does not respond to long-wavelength transverse disturbances, and the dispersion relation is given by
\be
 v_t^2 =\frac{S}{ (n_p+n_n^b)m}.
\ee
This differs from the usual relation for a one-component solid, $v_t^2=S/\rho$ by the replacement of the total mass density, $\rho$, by $(n_p+n_n^b)m$, the mass density of protons plus neutrons that move with the protons or, equivalently the total mass density less the density of superfluid neutrons, $\rho-m n_n^s$.

The dispersion relation for longitudinal waves has the form
\bea
\left(mv_l^2-n_n^s E_{nn}\right)\left(mv_l^2(n_p+n_n^b) - (\tilde K+4S/3 )\right)\nonumber \\ -L^2n_n^s -2mv_l^2n_n^bL-mv_l^2(n_n^b)^2 E_{nn}=0.
\eea
characteristic of two coupled modes \cite{sedrakian}. One mode is a sound wave in the neutron superfluid, which has a velocity $(n_n^s E_{nn}/m)^{1/2}$ and the second is a longitudinal lattice sound mode with velocity $( (\tilde K+4 S/3)/m(n_p+n_n^b)^{1/2}$.  The coupling between the two modes has three contributions, the $L^2$ term, which takes into account interactions between densities, the term proportional to $(n_n^b)^2$, which represents a current--current interaction, and a mixed term.  
In the crust of a neutron star at a density just above that for neutron drip, the velocity of the sound mode in the neutron superfluid is much less than that of the lattice mode, while closer to the inner boundary of the crust, the two velocity are similar and there is significant hybridization.  Detailed calculations will be reported elsewhere.

\section{The neutron superfluid density}
We may write Eq.\ (\ref{neutroncurrent}) for $\vv_p=0$ in the form $\jj_n=( n_n^s/m)\hbar\kk$.
The quantity $\chi_{jj}=n_n^s/m$ is the static, long-wavelength current--current correlation function for the neutrons, and it thus describes the change in the current density of neutrons when the crystal momentum $\hbar \kk$ of the superfluid is changed, the ions being stationary. Alternatively, one may write ${\bf g}_n=n_n^s (\hbar \kk)$.  The proton mass that enters the formalism in \S 3 is the bare mass, $m$, since for non-relativistic particles, the current density and the momentum density are related by the equation ${\bf g}_n=m\jj_n$.  One may write $\chi_{jj}=n_n^{\rm eff}/m_n^{\rm eff}$, where $n_n^{\rm eff}$ is an ``effective density of superfluid neutrons'' and $m_n^{\rm eff}$ a ``neutron effective mass''.  If one takes as 
$n_n^{\rm eff}$
the density of neutrons with single-particle energies greater than the mean field interaction energy outside nuclei, one is led to the conclusion that $m_n^{\rm eff}$
can be as large as $\sim 15m$~\cite{chameleffectivemass}. 
However, 
in the hydrodynamic equations, $n_n^{\rm eff}$ and $m_n^{\rm eff}$ will occur in the combination $n_n^{\rm eff}/m_n^{\rm eff}$, so it is irrelevant how one distributes changes in  $\chi_{jj}$ between a change in the effective mass and a change in the effective density of superfluid neutrons.  The quantity $n_n^s$ encapsulates all information about the superfluidity of neutrons that is needed for a hydrodynamic description: it includes effects due to the fact that many bands are filled, and that the effective mass of neutrons in  conduction bands may be different from the bare mass \cite{chameleffectivemass,chamel}.

\section{Concluding remarks}

In this paper we have indicated how to describe superfluidity of neutrons in the periodic lattice of ions in the inner crust of neutron stars.
We have also derived hydrodynamic equations to describe collective oscillations.  The key ingredients are the effective interactions 
$\partial \mu_i/\partial n_j, (i,j=n,p)$ between neutrons and protons, the shear elastic modulus $S$, and the ``neutron superfluid density''   
$n_n^s$.  The latter quantity describes how the neutron current density responds to a vector potential, and 
it is not simply related to the neutron density between nuclei. 
The related quantity  $n_n^b=n_n-n_n^s$    
determines both the effective mass of nuclei and the vector coupling between modes of the superfluid neutrons and lattice phonons. 
The fact that the same quantity occurs in these two contexts is a consequence of Galilean invariance of the combined system consisting of the lattice and neutrons. Applications of the formalism above will be reported elsewhere.

\section*{Acknowledgements}

This work was initiated at the Nordita workshop ``Neutron stars - The crust and beyond'' in September 2009 and continued at the Yukawa Institute in Kyoto at the workshop  ``New Frontiers in QCD 2010''.  We are grateful to the workshop organizers  and host institutions for support and hospitality. 
Support from FNRS (Belgium) and the ESF CompStar network are gratefully acknowledged. We thank Rishi Sharma and  Vincenzo Cirigliano for useful discussions.


\begin{thebibliography}{99}
\bibitem{neutronnormal}  We caution the reader that the neutron superfluid velocity ${\bf v}_n$ should not be confused with the velocity of the normal fluid for the two-fluid model for liquid helium-4.
\end{thebibliography}
\end{document}